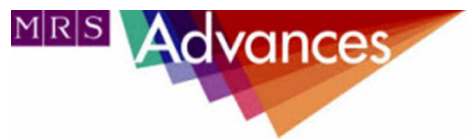

# Electronic Properties of CdS/CdTe Solar Cells as Influenced by the Choice of a Buffer Layer

# Electronic Properties of CdTe/CdS Solar Cells as Influenced by a Buffer Layer

Y. G. Fedorenko[1], J. D. Major[1], A. Pressman[1], L. Phillips[1], K. Durose[1]

[1]Stephenson Institute for Renewable Energy and Department of Physics, School of Physical Sciences, Chadwick Building, University of Liverpool, Liverpool L69 7ZF, UK

## ABSTRACT

We considered modification of the defect density of states in CdTe as influenced by a buffer layer in ZnO(ZnS, SnSe)/CdS/CdTe solar cells. Compared to the solar cells employing ZnO buffer layers, implementation of ZnSe and ZnS resulted in the lower net ionized acceptor concentration and the energy shift of the dominant deep trap levels to the midgap of CdTe. The results clearly indicated that the same defect was responsible for the inefficient doping and the formation of recombination centers in CdTe. This observation can be explained taking into account the effect of strain on the electronic properties of the grain boundary interface states in polycrystalline CdTe. In the conditions of strain, interaction of chlorine with the grain boundary point defects can be altered.

## INTRODUCTION

Owing to direct band gap of 1.5 eV and superior radiation hardness polycrystalline CdTe films are an essential component of many electronic devices such as detectors of high energy photons [1] and thin-film solar cells [2,3]. The charge carrier collection in CdTe solar cells has been found to crucially depend on the electronic properties of grain boundaries (GB) which are known to experience deformation and electrostatic fields due to charge trapping at the surface defects residing at the GB, therefore changing the band bending at the GBs and affecting the core GB current [4]. Electrical inactivation of the GB charge traps in CdTe has been predicted to proceed in the presence of Cl and Cu atoms which substitute for tellurium and cadmium, respectively, and co-passivate the Te core [5]. The performance of solar cells have been found to depend on the emission parameters of the defects, which are, in fact, the interface traps in the GBs [6,7]. The characteristics of the deep defect levels in the band gap of polycrystalline CdTe may be influenced by strain in the GBs, thus, suggesting high sensitivity of doping in CdTe to a particular deposition process and conditions of post-deposition anneals. CdTe is known as a "hard-to-dope" semiconductor [8] implying that the concept of a metal-oxide-semiconductor (MOS) structure implemented either by restricting the dimensions of a metal electrode [9] or modifying charge trapping in the interfacing CdS [10] can lead to the electric field screening and enable modulation of the surface band bending in CdTe. In the latter case, intentional doping of CdS with copper, which diffuses along the grain boundaries from the back contact of CdTe, or incorporation of a resistive ZnO buffer layer next to CdS in the solar cell structure have been suggested to increase the effective screening length in ZnO/CdS layers and result in the higher values of the open circuit voltage ($V_{OC}$) in CdS/CdTe solar cells. Therefore, electrostatic conditions at the CdS/CdTe junction can be influenced indirectly by using different buffer layers. In this work, we conducted investigation of the defect density of states (DOS) in the CdTe absorber as modified by the choice of a buffer layer in CdS/CdTe solar cells.

## EXPERIMENT

The solar cell CdS/CdTe structures were prepared identically except for introduction of different buffer layers, ZnO, ZnS, and ZnSe on TEC 7 glass covered with a conductive transparent $SnO_2$:F electrode [11]. The solar cell parameters for each series are shown in Table 1. No copper was introduced during the Au contact formation.

The solar cells were characterised by using capacitance-voltage ($C–V$) and ac admittance measurements carried out on a Solartron 1260 Impedance Analyzer equipped with a 1296 Dielectric Interface. The trap energy distributions $N_t(E)$ were deduced from the capacitance-frequency responses applying the analytical model for $p-i-n$ junction as proposed in work [12]. A Gaussian type of defect distribution was considered. To ensure that the carrier freeze-out effects do not contribute to the studied *ac* admittance responses, the activation energy $E_A$ of the series resistance $R_s$ was determined from the temperature dependent *IV*-curves by means of the differential method [13] as exemplified in Fig.1. The spatial distribution of the ionized acceptors in the depletion region of CdTe was extracted from the high-frequency capacitance-voltage ($CV$) curves.

| Buffer layer | Efficiency (%) | Fill Factor (%) | $J_{SC}$ (mA/cm$^2$) | $V_{OC}$ (V) |
|---|---|---|---|---|
| 120nm ZnO | 8.22 ± 1.70 | 59.25 ± 2.75 | 20.30 ± 1.67 | 0.675 ± 0.060 |
| 110nm ZnS | 0.85 ± 0.29 | 27.61 ± 3.39 | 6.03 ± 1.59 | 0.507 ± 0.024 |
| 100nm ZnSe | 0.98 ± 0.25 | 33.38 ± 1.92 | 6.22 ± 1.33 | 0.468 ± 0.022 |

**Table 1**. Averaged parameters of the solar cells used in this study.

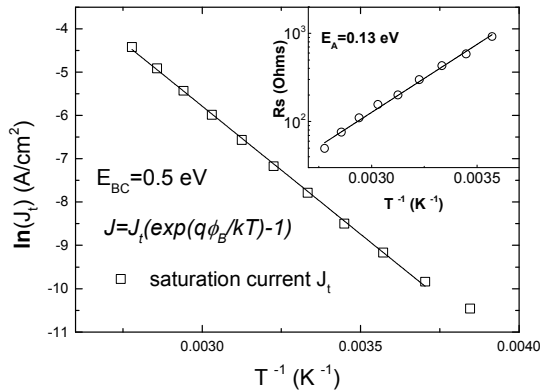

**Figure 1**. Arrhenius dependencies of the saturation current $J_t$ and the series resistance $R_s$ for ZnO/CdS/CdTe solar cells. $E_{BC}$ stands for the back contact barrier height.

## RESULTS

The Mott-Schottky plots for the samples grown on ZnO, ZnSe, and ZnS buffer layers and the distributions of the apparent carrier density in the absorber are shown in Fig. 2 (a, b), respectively. The cells comprising ZnS or ZnSe revealed lower values of the built-in voltage $V_{bi}$ being decreased to 0.54 V and 0.32 V, indicating a smaller barrier height at the heterojunction. The obtained values of $\rho$ throughout the CdTe thickness are very close in both samples, ZnS and ZnSe, and they are lower by approx. half of an order of magnitude than that of the cells grown on ZnO. The data presented in Fig. 2 (b) indicate that ZnS or ZnSe buffer layers inhibit the electrical activation by Cl atoms in CdTe. Similar non-uniform distributions of the net ionized impurity concentration were observed in the depletion region of the junction when the chlorine treatment was implemented in NaCl, KCl and $MnCl_2$ [11]. The lower doping efficiency has been ascribed to the cation charge state which differs from $2^+$ in chlorides other than $MgCl_2$ or $CdCl_2$. Later, inefficiency of anneals in Cl-containing ambient has been explained to originate from the higher dissociation energy of the cation−Cl bond in chlorides [14]. Importantly, the insufficient activation of acceptors altered the shape of the $CV$-carrier profiles. The latter is known to be determined by the spatial distribution of the charge traps in the space charge region (SCR) of a semiconductor, the charge state of the traps, and the emission time constants in respect to the time domain of $ac$ signal and the sweep voltage. The carrier profiles recorded for the solar cells concluded on ZnSe and ZnS buffer layers exhibited dips of similar magnitude at about $0.7W_D$. Partial recovery of the initial net ionized free carrier concentration is noticed on the profiles at the values of $(0.7 - 2.0) \times 10^{15}$ $cm^{-3}$ towards the full SCR width. As a first approximation, the difference between the profiles can be accounted for the presence of intrinsic point defects which may contribute to the compensation in semi-insulating CdTe. Among the studied deep trap levels in single-crystal CdTe, the recombination active centre is a deep acceptor complex with a trap level at $E_V + 0.76$ eV in undoped and doped CdTe [15]. In polycrystalline CdTe, the identification of a particular point defect giving rise to the recombination losses in solar cells may be complicated by superimposed contributions from the interface states in the GBs [7].

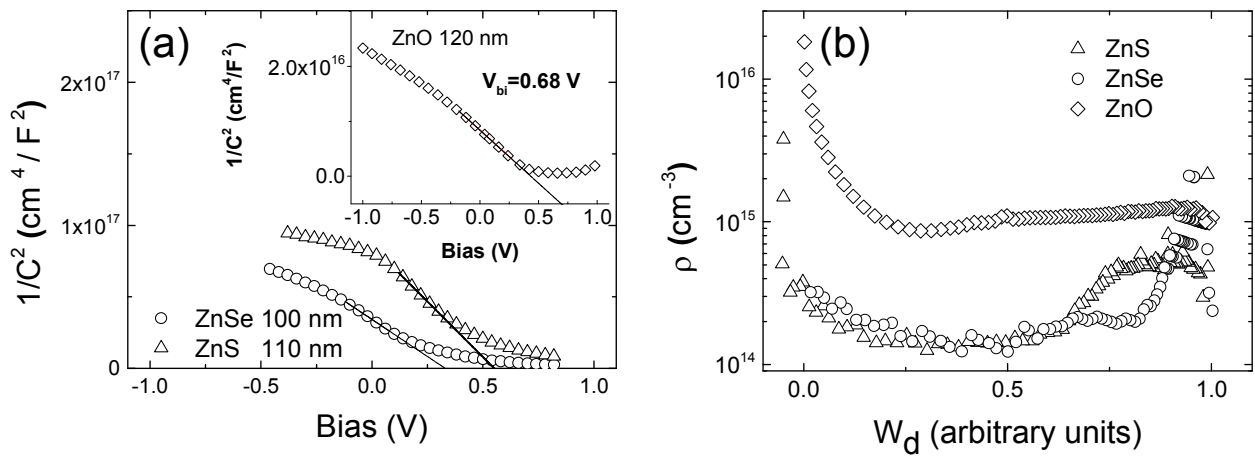

**Figure 2**. (a) $1/C^2–V$ plots for samples with different buffer layers. The data were measured at T =300 K and frequency of 10 kHz. The inset, Fig. 2(a), shows the Mott–Schottky plot for the ZnO/CdS/CdTe solar cell. The straight black lines are the fitting curves to estimate the built-in potential $V_{bi}$; (b) The apparent doping density profiles $\rho(W_D)$ obtained from $C–V$ characteristics in Fig.2(a).

Figure 3(b-d) shows the energy distributions of the charge traps in the CdTe/CdS junction using the $E_A$ values extracted from the Arrhenius plots in Fig. 3(a). Only one deep defect level is observed in the studied samples in accordance to earlier findings on CdTe/CdS solar cells subjected to the NP etch prior to evaporation of Au contacts [6]. A significant energy shift of the activation energy from 0.38 eV (Fig. 3(b)), to 0.5 eV (Fig. 3 (c)), and to 0.75-0.76 eV (Fig. 3

(d)), was observed for the CdTe/CdS cells grown on ZnO, ZnSe, and ZnS buffer layers, respectively. The Gaussian fit of the DOS energy distribution provides the values of the broadening parameter $\delta$, which was found to increase from 0.26 meV to 100 meV in the temperature range from 180 K to 400 K for the DOS displayed in Figs. 3(b, c). The broadening of the DOS may be caused by the locally non-uniform distribution of charged defects and defect clusters resulting in potential fluctuations, contributing to Shockley-Read-Hall recombination, and decreasing the open circuit voltage and the fill factor of solar cells.

Figure 4 compiles *IV* characteristics taken on the ZnO/CdTe/CdS cells. The charge carrier transport mechanism at both the forward, Fig 4(a), and the reverse, Fig. 4(b), bias is space-charge-limited conduction (SCLC) as evidenced by the slope of the *log(I)-log(V)* curves increasing from 1 to 1.4-1.6 that corresponds to a monoenergetic trap level, which can be expected to be shallow in the CdTe/CdS junction since a steep increase in the trap filling charge regime is not observed. The relatively high ohmic current at low voltages is most probably caused by high concentration of impurities in the utilized semiconductors. The trap density $N_t$ can be estimated by using the expression for the critical voltage $V_T = qN_td^2/2\varepsilon\varepsilon_0$, were $q$ is the elemental charge, $d$ is the film thickness, $\varepsilon$ is the film permittivity, and $\varepsilon_0$ is the permittivity of the free space. Given that the relative permittivity for CdTe is 10.16, the trap density $N_t$ can be estimated as $\sim 7 \cdot 10^{15}$ cm$^{-3}$ and $\sim 4.2 \cdot 10^{16}$ cm$^{-3}$ for positive and negative polarity of the applied bias, respectively, that is much higher than the density $N_{it}$ of the traps inferred from the Gaussian fit of the trap distributions in energy. The asymmetry in *IV* curves implies different rates of the carrier generation and recombination in the space-charge region of CdTe. Though the asymmetric *IV* curves may indicate the presence of Schottky barrier as it has been typically observed in metal-semiconductor-metal structures with dissimilar work functions of electrodes, in the particular case of CdTe solar cells, the blocking action of the TCO/buffer layers/CdS interfaces is unlikely. The charge carrier transport in polycrystalline semiconductors employed in photovoltaics is often supported by tunnelling processes [16]. Having obtained the trap-free space-charge-limited current density on the order of $10^{-5}$ A/cm$^2$ from the 300K *IV* curve in Fig. 4(a) the charge carrier mobility $\mu$ of $\sim 0.1$ cm$^2$/V·s can be deduced by using the expression $J=9/8 \cdot \theta_0 \cdot \varepsilon_0 \cdot \varepsilon \cdot \mu \cdot V^2/d^3$, where $E$ is the electric field across the device, $d$ is the thickness of the absorber layer, and $\theta_0$ is the fraction of the free carriers [17]. The obtained mobility values are in good agreement with those measured by using the time-of-flight method in work [18]. Assuming the charge carrier density in p-type CdTe is lower than that in CdS, the injected charge density is mostly represented by the electron component, and the determined mobility describes the electron current.

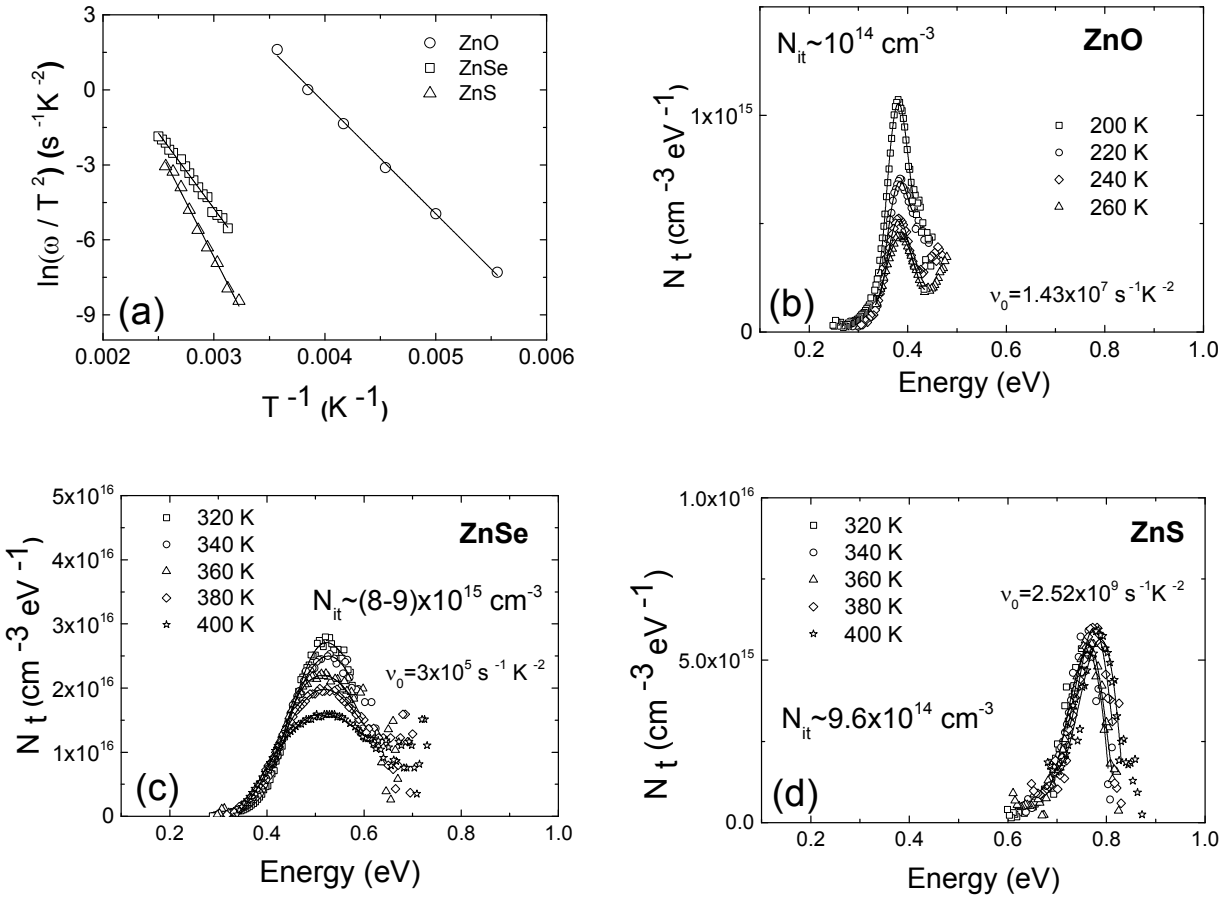

**Figure 3**. The Arrhenius plots (a) and the density of states in the CdS/CdTe solar cells deposited on different buffer layers: (b) ZnO, (c) ZnSe, (d) ZnS. The emission rates were determined from the Arrhenius plots taken at zero bias.

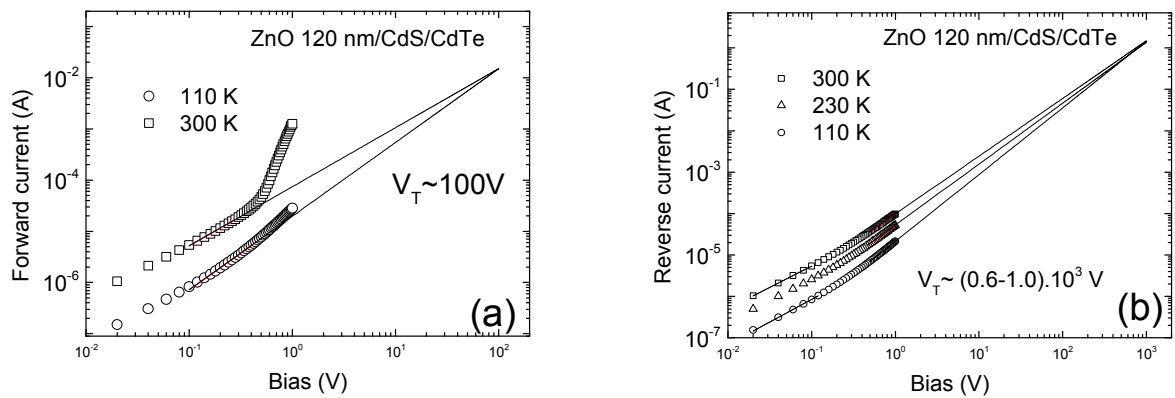

**Figure 4**. Forward (a) and reverse (b) current-voltage characteristics of the CdTe/CdS junctions deposited on ZnO. The contact area is 0.24 cm$^2$.

**CONCLUSIONS**

The deep trap level formation in the solar cells comprising ZnS or ZnSe buffer layers is accompanied by a reduction of the doping density in CdTe implying that the electrical activation of acceptors by chlorine is impeded. This suggests that chlorine can modify kinetics of point defects during anneals of CdTe and redistribute $V_{Cd}$ –related defects, similar to laser anneals, which generate a gradient of the $V_{Cd}$ defects in CdTe [19,20]. We suggest that a higher density of the deep charge traps in CdTe could be ascribed to the $V_{Cd}$ defect clustering in polycrystalline CdTe:Cl, although the $V_{Cd}$ clustering in bulk crystals and in the GBs of polycrystalline samples may differ in kinetics. The defect clustering might occur when the chlorine distribution is locally non-uniform as a result of strain in the GB. Therefore, passivation of the Te-rich GBs may be impeded. Our results clearly indicate that the same defect is responsible for doping and the formation of the charge trapping centers in CdTe. In general, grain boundaries have been implicated in current leakage and electric breakdown of a variety of electronic devices including thin film solar cells. Being segregated in grain boundaries, $V_{Cd}$ related defects may play a decisive role in electron tunnelling and charge trapping. Further, it is observed that the majority of electrically active defects in the studied polycrystalline CdTe are the band tail-like defects, which influence conductivity. The charge carrier transport in the studied CdS/CdTe heterojunctions does not obey the Sah–Noyce–Shockley theory, but can be described within the theory of space-charge limited conduction.

## ACKNOWLEDGMENTS

This work was supported by EPSRC PPR10314.